# Digital-Analog Transmission based Emergency Semantic Communications

Yuzhou Fu, *Student Member, IEEE*, Wenchi Cheng, *Senior Member, IEEE*, Jingqing Wang, *Member, IEEE*, Liuguo Yin, *Senior Member, IEEE*, and Wei Zhang, *Fellow, IEEE*

*Abstract*—Emergency Wireless Communication (EWC) networks adopt the User Datagram Protocol (UDP) to transmit scene images in real time for quickly assessing the extent of the damage. However, existing UDP-based EWC exhibits suboptimal performance under poor channel conditions since UDP lacks an Automatic Repeat reQuest (ARQ) mechanism. In addition, future EWC systems must not only enhance human decision-making during emergency response operations but also support Artificial Intelligence (AI)-driven approaches to improve rescue efficiency. The Deep Learning-based Semantic Communication (DL-based SemCom) emerges as a robust, efficient, and task-oriented transmission scheme, suitable for deployment in UDP-based EWC. However, the performance of the existing DL-based SemCom is significantly influenced by the integrated Neural Network (NN) models. Due to the constraints in hardware capabilities and transmission resources, the EWC transmitter is unable to integrate sufficiently powerful NN model, thereby failing to achieve ideal performance under EWC scene. This limitation highlights the urgent need for further exploration of robust and efficient DL-based SemCom framework for the UDP-based EWC networks. For EWC scene, we propose a performance-constrained semantic coding model, which considers the effects of the semantic noise and the channel noise. Then, we derive Cramér-Rao lower bound of the proposed semantic coding model, as guidance for the design of semantic codec to enhance its adaptability to semantic noise as well as channel noise. To further improve the system performance, we propose Digital-Analog transmission based Emergency Semantic Communication (DA-ESemCom) framework, which integrates the analog DL-based semantic coding and the digital Distributed Source Coding (DSC) schemes to leverage their respective advantages. The simulation results show that the proposed DA-ESemCom framework outperforms the classical Separated Source-Channel Coding (SSCC) and other DL-based Joint Source-Channel Coding (DL-based JSCC) schemes in terms of fidelity and detection performances.

## I. Introduction

Unforeseen disasters, whether of natural origin or caused by human activities, such as Mexico earthquake in 2017, Brazil's forest fires in 2024, and Parker fire in California, etc., have profound and devastating effects on human lives and communities. In the disaster area, the communication infrastructures are severe damaged. At the same time, the victims affected by the disaster and the rapidly deployed devices are trying to convey the information about their circumstances, generating high-bandwidth data in the form of images, while the surviving communication infrastructure fails to support the dramatic increase in data [1]. After a disaster, Emergency Wireless Communication (EWC) [2], [3] that establishes high reliability and low latency wireless transmission during emergency situations, plays a crucial role for quickly assessing an extent of the damage and efficiently performing rescues. However, the existing EWC framework is difficult to provide robust data transmission under poor channel conditions, thus effectively supporting the rescue actions.

For achieving real time data transmission, the EWC networks forward the data by using User Datagram Protocol (UDP) [4], [5], which lacks Automatic Repeat reQuest (ARQ) mechanism. The existing UDP-based EWC relies on classic Shannon's transmission framework and adopts Separated Source-Channel Coding (SSCC) scheme, which exhibits suboptimal performance under poor channel conditions. In the EWC networks, the Unmanned Aerial Vehicles (UAVs) [6] and Emergency Communication Vehicles (ECVs) [7] are often used as the relay or temporary Base Station (BS) to forward the arriving data to the post-disaster area. Under constrained total energy, the energy for wireless transmission of the UAVs and ECVs are very limited since they must support not only wireless communication but also other operational functions, including flight and auxiliary device usage. Furthermore, the wireless link between the transmitter at the disaster area and the receiver at the rear area undergoes large scale fading. Thus, the EWC often experiences low Signal-Noise-Ratio (SNR) channel condition, which poses a significant challenge for achieving robust UDP-based EWC network.

Under the low SNR conditions, the DL-based SemCom demonstrates superior robustness compared to traditional reliable communication due to its enhanced semantic recovery capabilities and inherent adaptability to channel noise [8]. In particular, the authors of [9] proposed DL enabled end-to-end semantic communication framework, which can extract the semantic feature of the data and exactly recovers the meaning of original data with the error transmission. The authors of [10] proposed DL-based JSCC scheme to learn noise resilient encoded symbols for image transmission over noise channels, thus achieving robustness to poor channel conditions. In [11], the authors proposed a robust semantic communication frame-

Yuzhou Fu, Wenchi Cheng, and Jingqing Wang are with the School of Telecommunications Engineering, Xidian University, Xi'an, 710071, China (e-mails: fyzhouxd@stu.xidian.edu.cn; wccheng@xidian.edu.cn; jqwangxd@xidian.edu.cn).

Liuguo Yin is with the Beijing National Research Center for Information Science and Technology, Tsinghua University, Beijing, 100084, China (e-mail: yinlg@tsinghua.edu.cn).

Wei Zhang is with the School of Electrical Engineering and Telecommunications, the University of New South Wales, Sydney, Australia (e-mail: w.zhang@unsw.edu.au).

work and designed the masked Vector Quantized-Variational AutoEncoder (VQ-VAE) to combat the semantic noise and channel noise. For robust forward data, the authors of [12] proposed the semantic-forward relaying scheme, which can reduce forwarding payload, and also improve the network robustness against intra-link errors. Therefore, the DL-based SemCom is a potential technology and can be used for UDP-based EWC to improve its robustness at the low SNR channel conditions. However, the existing DL-based SemCom frameworks can achieve good performance at low SNR region, but its reachable performance is significantly influenced by the Neural Network (NN) model integrated into it. Due to the limited transmission and hardware resources [13], the NN model suffers serious performance degradation in the emergency communication environment. As a result, the performance of the DL-based SemCom is severely limited even under good channel conditions. In such emergency communication scene, it is highly demanded for developing a performance-enhancing DL-based SemCom under the good channel condition.

Apart from the development of robust emergency communication, the exploding amount of data to be forwarded in the disaster areas results in the communication congestion, which imposes challenges for efficient data processing and transmission schemes. Also, Artificial Intelligence (AI)-driven methods, characterized by reduced human intervention, faster response times, cost-effectiveness, and broader surveillance coverage in disaster management and assistance, are increasingly being utilized or proposed for future improvements in disaster management [14], [15]. For reducing the operation delay of the AI-driven method [16], the EWC not only needs to provide efficient data transmission by further compressing original data, but also tightly couples with the AI-driven method. Compared to the classical SSCC scheme, the DL-based SemCom aims to adopt semantic coding scheme, which can more efficiently compress the source sequences by merging sequences of the same semantics [8]. Also, the semantic coding scheme is able to only transmit the necessary information according to the requirement of AI-driven applications at the receiver [17]. The irrelevant parts will be abandoned or de-emphasized, thus saving transmission resources. For example, the authors of [18] studied the task-oriented communication algorithm for image classification task offloading in the aerial systems and proposed a joint semantic extraction and compression model for achieve efficient semantic transmission under different channel conditions. The authors of [19] explored personalized saliency-based task-oriented semantic communication in UAV image sensing scenarios and presented an energy-efficient task-oriented semantic communication framework for the image retrieval. Thus, the semantic coding scheme should be integrated into EWC, thus achieving efficient data transmission and well support the AI-driven method. However, the existing semantic coding frameworks for serving the AI-driven method often exhibit suboptimal performance in terms of reconstruction quality, as they aim to ensure semantic reasoning rather than data fidelity.To address various demands of emergency rescue, the reconstructed data of the semantic coding-based EWC not only needs to support the AI-driven method, but also facilitate human decision-making during emergency response operations. Different from the widely used wireless networks, the EWC network is difficult to ingrate multiple NN model for differentiated semantic coding according to the intention of the receiver. Thus, we need to further investigate a common DL-based SemCom framework, which can achieve good performance on the reconstruction quality as well as AI-driven method.

Motivated by the development of DL-based SemCom for EWC environment, we investigate a novel Emergency Semantic Communication (ESemCom) framework, which adopts the semantic coding scheme to forward the data from the transmitter at the disaster area to the receiver at the rear area. We consider a point-to-point emergency communication scenario, where the transmitter has constraints in hardware capabilities and energy availability. In particular, we propose performance-constrained semantic coding model with the effects of the semantic noise and the channel noise, where the semantic noise is caused by the errors of the semantic extraction and the semantic transmission. Then, we derive Cramér-Rao lower bound of the proposed performance-constrained semantic coding model and analyze its performance bottleneck. To improve the system performance of the ESemCom, we propose Digital-Analog transmission based Emergency Semantic Communication (DA-ESemCom) framework, which integrates the analog semantic coding and the digital traditional Distributed Source Coding (DSC) schemes, thus leveraging their advantages on the noise-robustness and data fidelity. In addition, we develop a DA transmission-based coding scheme for the proposed DA-ESemCom framework with limited power and bandwidth resources. Numerical results show that our proposed DA-ESemCom framework performs better than the classical SSCC and DL-based JSCC schemes in terms of reconstruction quality and detection performance.

The rest of this paper is organized as follows. Section II proposes performance-constrained semantic coding model for the ESemCom. In Section III, we integrate the digital-analog transmission framework into the ESemCom to further improve system performance. Then, in Section IV, the DA transmission-based coding scheme is proposed for our proposed ESemCom framework. Performance analyses are given in Section V. Finally, the conclusions are drawn in Section VI.

## II. PROPOSED PERFORMANCE-CONSTRAINED SEMANTIC CODING MODEL FOR EWC SCENE

In this section, we propose performance-constrained semantic coding model for point-to-point EWC scene, as shown in Fig. 1. In particular, the mobile EWC device within disaster area is not only equipped with the sensing devices to collect scene image, but also can be used as the transmitter to forward the data [20]. Also, the transmitter can only deploy performance-constrained semantic encoder for image transmission due to limited hardware resources. In practice, the image encoding and transmission adopt the image patch-based processing approach, where an image is split into $I$ non-overlapping and normalized image patches for subsequent encoding [21]. The EWC adopts the UDP-based data transmission, where the transmission block will be distorted by

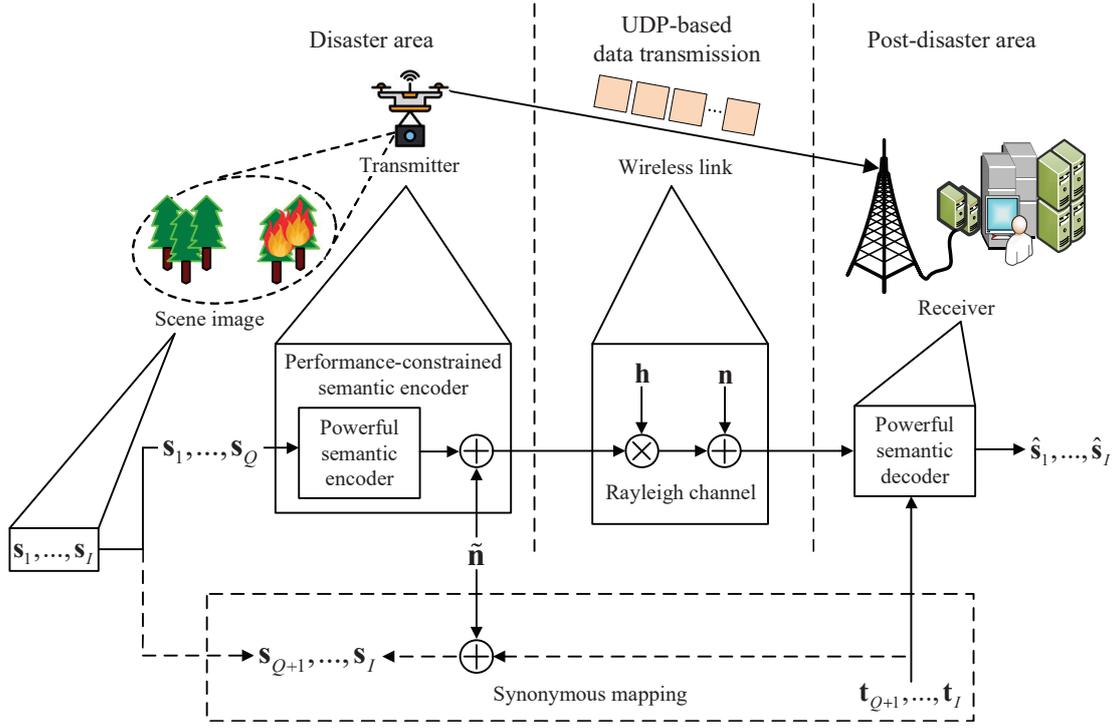

Fig. 1: The proposed performance-constrained semantic coding model for EWC scene.

Rayleigh fading as well as Additive White Gaussian Noise (AWGN). In the post-disaster area, the receiver with sufficient computing resources adopts powerful semantic decoder, which works under full-performance to perform robust image recovery.

*A. The Architecture of Proposed Semantic Coding Model*

Figure 1 shows the proposed performance-constrained semantic coding model, where the semantic encoder is modeled as a powerful semantic encoder affected by the semantic noise. The powerful semantic encoder can map full set of the image patch to the semantic features with infinite constellation space. However, the semantic encoder deploying in the EWC only can achieve finite constellation mapping caused by constrained hardware resources [13]. Also, due to severely limited transmission bandwidth in the EWC scene, the semantic encoder is only mapped a subset of the image patches to the encoded symbols for the wireless transmission. The rest of the image patches are processed by synonymous mapping without sending to the receiver. With the synonymous mapping [22], a set of the image patches, which indicates similar meaning, such as sky, tree, and ground, can use a shared sequence to represent and can achieve a satisfactory recovery at powerful semantic decoder. Thus, we can consider that the semantic noise is produced by finite constellation mapping and synonymous mapping error.

*B. Semantic Coding and Wireless Transmission Model*

For following modeling and analysis, the image patches is modeled as a set of Gaussian sources, which is defined as $\mathbf{S} = \{\mathbf{s}_i\}_{i=1}^{I}$. In particular, $\mathbf{s}_i$ with $i = 1, 2, ..., Q$ is encoded as encoded symbols for wireless transmission. Also, $\mathbf{s}_i$ with $i = Q+1, Q+2..., I$ is processed by synonymous mapping without wireless transmission. Let be $f_E(\mathbf{s}_i)$ the mapping function of the semantic encoder to encode the source $\mathbf{s}_i$. Then, the set of received encoded symbol is defined as $\mathbf{Y} = \{\mathbf{y}_i\}_{i=1}^{Q}$, where $\mathbf{y}_i \in \mathbb{C}^{L \times 1}$ is $i$th received symbol and can be expressed as follows:

$$\mathbf{y}_i = (f_E(\mathbf{s}_i) + \widetilde{\mathbf{n}}) h_i + \mathbf{n}, \quad (1)$$

where $i = 1, 2, ...Q$ and $\widetilde{\mathbf{n}} \in \mathbb{C}^{Q \times 1} \sim \mathcal{CN}(0, \sigma_{\widetilde{n}}^2)$ denotes the semantic noise, $h_i \in \mathbb{C}^{1 \times 1} \sim \mathcal{CN}(0, 1)$ is the channel fading gain, and $\mathbf{n} \in \mathbb{C}^{Q \times 1} \sim \mathcal{CN}(0, \sigma_n^2)$ is the AWGM.

For synonymous mapping, the semantic decoder learns tokens for synonymous mapping. Those tokens aim to maximize the fidelity of the estimated sources. The token set is defined as $\mathbf{T} = \{\mathbf{t}_i\}_{i=Q+1}^{I}$, where $\mathbf{t}_i \sim \mathcal{CN}(0, \sigma_{t,i}^2)$ is $i$th token corresponding the dropped source $\mathbf{s}_i$. In particular, $\mathbf{t}_i$ is conditionally independent given $\mathbf{s}_i$ since the value of $\mathbf{t}_i$ is fixed after deep learning training. Also, the synonymous mapping exists inevitably fitting error between the learned tokens and the specific sources, where the fitting error can be modeled as the semantic noise. Then, $\{\mathbf{s}_i\}_{i=Q+1}^{I}$ can be given as follows:

$$\mathbf{s}_i = \mathbf{t}_i + \widetilde{\mathbf{n}}, \quad (2)$$

where $i = Q+1, Q+2, ..., I$.

*C. Distortion Analysis of proposed Performance-constrained Semantic Coding Model*

In the following, we analyze the overall distortion of proposed performance-constrained semantic coding model. The output of the semantic decoder, denoted by $\hat{\mathbf{S}} = \{\hat{\mathbf{s}}_i\}_{i=1}^I$, and can be given by
$$\hat{\mathbf{S}} = f_D(\mathbf{S}|\mathbf{Y}, \mathbf{T}), \quad (3)$$

where $f_D(\cdot)$ represents the semantic decoder. The overall distortion, denoted by $D_{\text{overall}}$, is a combination of semantic transmission distortion and synonymous mapping distortion. With unbiased mapping of the semantic decoder, the semantic transmission distortion is the error between ideal semantic extraction and the channel output. Also, the synonymous mapping distortion denotes the gap between the original sources and the learned tokens. Base on (1), (2), and (3), $D_{\text{overall}}$ can be formulated as follows:

$$D_{\text{overall}} = \underbrace{\sum_{i=1}^Q \mathbb{E}\left[(f_E(\mathbf{s}_i) - \mathbf{y}_i)^2\right]}_{\text{semantic transmission distortion}} + \underbrace{\sum_{i=Q+1}^I \mathbb{E}\left[(\mathbf{s}_i - \mathbf{t}_i)^2\right]}_{\text{synonymous mapping distortion}}. \quad (4)$$

The Cramér-Rao lower bound (CRLB) [23] can give a lower bound on the distortion for the unbiased mapping of the semantic decoder (3). Also, the CRLB can provide useful the design of good semantic encoder mappings [24]. The CRLB of the overall distortion, denoted by $\overline{D}_{\text{overall}}$, can be expressed as follows:

$$D_{\text{overall}} \geq \overline{D}_{\text{overall}}$$
$$= \underbrace{\sum_{i=1}^Q \left\{-\mathbb{E}\left[\frac{\partial^2 \ln p(\mathbf{y}_i|\mathbf{s}_i, h_i)}{\partial \mathbf{s}_i^2}\right]\right\}^{-1}}_{\text{CRLB of semantic transmission distortion}}$$
$$+ \underbrace{\sum_{i=Q+1}^I \left\{-\mathbb{E}\left[\frac{\partial^2 \ln p(\mathbf{t}_i|\mathbf{s}_i)}{\partial \mathbf{s}_i^2}\right]\right\}^{-1}}_{\text{CRLB of synonymous mapping distortion}}. \quad (5)$$

Based on (1), we can obtain $h_i \widetilde{n} + n \sim \mathcal{CN}(0, \frac{\sigma_{\widetilde{n}}^2}{1+\sigma_{\widetilde{n}}^2} + \sigma_n^2)$, and $p(\mathbf{y}_i|\mathbf{s}_i, h_i)$ is given as follows:

$$p(\mathbf{y}_i|\mathbf{s}_i, h_i) = \left[\frac{1}{2\pi \times \left(\frac{\sigma_{\widetilde{n}}^2}{1+\sigma_{\widetilde{n}}^2} + \sigma_n^2\right)}\right]^{L/2}$$
$$\times \exp\left\{-\frac{1}{2\left(\frac{\sigma_{\widetilde{n}}^2}{1+\sigma_{\widetilde{n}}^2} + \sigma_n^2\right)^2} \sum_{l=1}^L [f_E(s_{i,l})h_i - y_{i,l}]^2\right\}, \quad (6)$$

where $s_{i,l}$ is $l$-th normalized pixel value of $\mathbf{s}_i$, $L$ is the length of $\mathbf{s}_i$, $y_{i,l}$ denotes $l$th random sample of $i$th channel output, and $f_E(s_{i,l})$ represents $l$th value of $i$th semantic encoder output. Then, the variance of $\mathbf{s}_i$ can be given as follows [25]:

$$\sigma_{s,i}^2 = \frac{\sum_{l=1}^L (s_{i,l} - \mu_i)}{L}, \quad (7)$$

where $\mu_i$ represents the mean value of $\mathbf{s}_i$ and $\sigma_{s,i}^2$ reflects the dispersion degree or texture complexity of $i$-th image patch.

Since $\mathbf{T}$ is conditionally independent given $\mathbf{X}$, $p(\mathbf{s}_i|\mathbf{t}_i)$ can be given as follows:

$$p(\mathbf{t}_i|\mathbf{s}_i) = \frac{p(\mathbf{s}_i, \mathbf{t}_i)}{p(\mathbf{s}_i)} = \frac{p(\mathbf{t}_i)p(\mathbf{n})}{p(\mathbf{s}_i)} = \left(\frac{\sigma_{s,i}^2}{2\pi\sigma_{t,i}^2\sigma_{\widetilde{n}}^2}\right)^{L/2}$$
$$\times \exp\left[\sum_{l=1}^L \left(-\frac{t_{i,l}^2}{2\sigma_{t,i}^2} - \frac{\widetilde{n}_l^2}{2\sigma_{\widetilde{n}}^2} + \frac{\widetilde{n}_l^2}{2\sigma_{\widetilde{n}}^2}\right)\right]$$
$$= \left(\frac{\sigma_{s,i}^2}{2\pi\sigma_{t,i}^2\sigma_{\widetilde{n}}^2}\right)^{L/2}$$
$$\times \exp\left[-\frac{\sigma_{s,i}^2}{2\sigma_{t,i}^2\sigma_{\widetilde{n}}^2} \sum_{l=1}^L \left(t_{i,l} - \frac{\sigma_{t,i}}{\sigma_{s,i}}s_{i,l}\right)^2\right], \quad (8)$$

where $t_{i,l}$ is $l$th random sample of $i$th token. Then, by taking the logarithm and thereafter deriving the second derivative with respect to $\mathbf{s}_i$, (6) is transformed as follows:

$$\frac{\partial^2 \ln p(\mathbf{y}_i|\mathbf{s}_i, h_i)}{\partial^2 \mathbf{s}_i} = \frac{1}{\left(\frac{\sigma_{\widetilde{n}}^2}{1+\sigma_{\widetilde{n}}^2} + \sigma_n^2\right)}$$
$$\times \sum_{l=1}^L \left\{\left[\frac{\partial h_i f_E(s_{i,l})}{\partial s_{i,l}}\right]^2 - [y_{i,l} - h_i f_E(s_{i,l})]\frac{\partial^2 h_i f_E(s_{i,l})}{\partial^2 s_{i,l}}\right\}, \quad (9)$$

then (8) is also transformed as follows:

$$\frac{\partial^2 \ln p(\mathbf{t}_i|\mathbf{s}_i)}{\partial^2 \mathbf{s}_i} = -L\frac{\sigma_{t,i}^2}{\sigma_{s,i}^2\sigma_{\widetilde{n}}^2}. \quad (10)$$

We can calculate the mathematical expectation of (9) as follows:

$$\mathbb{E}\left[\frac{\partial^2 \ln p(\mathbf{y}_i|\mathbf{s}_i, h_i)}{\partial^2 \mathbf{s}_i}\right]$$
$$= -\frac{1}{\left(\frac{\sigma_{\widetilde{n}}^2}{1+\sigma_{\widetilde{n}}^2} + \sigma_n^2\right)} \times \sum_{l=1}^L \left\{\mathbb{E}\left[\left(\frac{\partial h_i f_E(s_{i,l})}{\partial s_{i,l}}\right)^2\right]\right.$$
$$\left. - \mathbb{E}\left[(y_{i,l} - f_E(s_{i,l}))\frac{\partial^2 h_i f_E(s_{i,l})}{\partial^2 s_{i,l}}\right]\right\}$$
$$= -\frac{1}{\left(\frac{\sigma_{\widetilde{n}}^2}{1+\sigma_{\widetilde{n}}^2} + \sigma_n^2\right)} \times \sum_{l=1}^L \mathbb{E}\left[\left(\frac{\partial f_E(s_{i,l})}{\partial s_{i,l}}\right)^2\right]. \quad (11)$$

Also, the mathematical expectation of (10) can be obtained as follows:

$$\mathbb{E}\left[\frac{\partial^2 \ln p(\mathbf{t}_i|\mathbf{s}_i)}{\partial^2 \mathbf{s}_i}\right] = -L\frac{\sigma_{t,i}^2}{\sigma_{s,i}^2\sigma_{\widetilde{n}}^2}. \quad (12)$$

Based on (5), the CRLB of the overall distortion can be expressed as follows:

$$\overline{D}_{\text{overall}} = \sum_{i=1}^Q \frac{1}{\frac{1}{\left(\frac{\sigma_{\widetilde{n}}^2}{1+\sigma_{\widetilde{n}}^2}+\sigma_n^2\right)}\sum_{l=1}^L \mathbb{E}\left[\left(\frac{\partial f_E(s_{i,l})}{\partial s_{i,l}}\right)^2\right]}$$
$$+ \sum_{i=Q+1}^I \frac{1}{L\frac{\sigma_{t,i}^2}{\sigma_{s,i}^2\sigma_{\widetilde{n}}^2}}. \quad (13)$$

The correlation coefficient between $\mathbf{s}_i$ and $\mathbf{t}_i$, denoted by $\rho_i$, and is given by

$$\begin{aligned}\rho_i &= \frac{\mathbb{E}(\mathbf{s}_i\mathbf{t}_i) - \mathbb{E}(\mathbf{s}_i)\mathbb{E}(\mathbf{t}_i)}{\sigma_{s,i}\sigma_{t,i}} \\ &= \frac{\mathbb{E}\left[\mathbf{t}_i(\mathbf{t}_i+\widetilde{\mathbf{n}})\right] - \mathbb{E}(\mathbf{t}_i+\widetilde{\mathbf{n}})\mathbb{E}(\mathbf{t}_i)}{\sigma_{s,i}\sigma_{t,i}} \\ &= \frac{\mathbb{E}\left[(\mathbf{t}_i)^2\right] - [\mathbb{E}(\mathbf{t}_i)]^2}{\sigma_{s,i}\sigma_{t,i}} \\ &= \frac{\sigma_{t,i}^2}{\sigma_{s,i}\sigma_{t,i}} = \frac{\sigma_{t,i}}{\sigma_{s,i}}.\end{aligned} \quad (14)$$

The fitting coefficient reflects the effect of fitting one random variable with another, and it is equal to the square of its correlation coefficient. Then, the fitting coefficient between $\mathbf{s}_i$ and $\mathbf{t}_i$, denoted by $\rho_i^2 \in [0,1]$, and is given as follows:

$$\rho_i^2 = \left(\frac{\sigma_{t,i}}{\sigma_{s,i}}\right)^2, \quad (15)$$

thus, (13) can be rewritten as follows:

$$\overline{D}_{\text{overall}} = \sum_{i=1}^{Q} \frac{\sigma_{\widetilde{n}}^2(1+\sigma_n^2) + \sigma_n^2}{(1+\sigma_{\widetilde{n}}^2)\sum_{l=1}^{L}\mathbb{E}\left[\left(\frac{\partial f_E(s_{i,l})}{\partial s_{i,l}}\right)^2\right]} + \sum_{i=Q+1}^{I} \frac{\sigma_{\widetilde{n}}^2}{L\rho_i^2}. \quad (16)$$

From (16), $\rho_i^2$ should be as high as possible to limit the effect of the semantic noise, thus minimizing the synonymous mapping distortion. For the synonymous mapping, the transmitter should select the source, which is with similar distribution with the token. Also, (16) can be used to design the optimization objective to train the semantic codec for learning the optimal token and coded representations, thus enhancing its semantic recovery capabilities and inherent adaptability to semantic noise as well as channel noise.

*D. The Optimization Objective and Training Algorithm*

Based on (16), we propose an optimization objective to train the semantic codec. For the end-to-end semantic communication, the optimization objective consists of two loss functions: the overall distortion loss as (16) and reconstruction loss. Specifically, the overall distortion loss is used to train the semantic encoder, which can learn the best code mapping and synonymous mapping to effectively preserve the meaning of the sources. Also, the reconstruction loss focuses on training the semantic decoder to maximize fidelity as the goal to restore the data. Therefore, the Mean Square Error (MSE) between the sources and the estimated outputs is used the reconstruction loss function, which denoted by $\mathcal{L}_{\text{rec}}$ and is given as follows:

$$\mathcal{L}_{\text{rec}} = \frac{1}{I}\sum_{i=1}^{I} ||\mathbf{s}_i - \hat{\mathbf{s}}_i||_2^2. \quad (17)$$

Then, the optimization objective can be formulated as follows:

$$\min_{\boldsymbol{\theta},\boldsymbol{\beta}} \left(\overline{D}_{\text{overall}} + \mathcal{L}_{\text{rec}}\right), \quad (18)$$

where $\boldsymbol{\theta}$ and $\boldsymbol{\beta}$ are trainable neural network parameters of the semantic encoder and the semantic decoder, respectively.

Then, we develop a training algorithm to ensure the proposed semantic coding with the best performance after training. The training of the semantic coding adopts the end-to-end manner training with random masking scheme. Consequently, the semantic encoder and the semantic decoder can learn the optimal code mapping representations to achieve robust semantic transmission over the noise channels. During iterative training, we adopts a pre-trained ideal semantic encoder to estimate the semantic noise with maximum likelihood estimation. The pre-trained ideal semantic encoder is defined as $f_{\text{E}}^{\text{Ieal}}(\cdot)$, which adopts semantic encoding with all sources and full symbol precision. The masking ratio, denoted by $\eta = \frac{Q}{I}$, refers to the ratio of the number of dropped sources and full sources. The learning rate, training epoch, and trained SNR are defined as $\alpha$, $\varepsilon$, and $\gamma$. The training algorithm is summarized in **Algorithm 1**. The proposed optimization

---

**Algorithm 1** Semantic Coding Training Algorithm

1: **Input:** Dataset $\mathcal{D}$, $f_{\text{E}}^{\text{Ieal}}(\cdot)$, $\eta = 0.7$, $\varepsilon = 100$, $\alpha = 1e - 6$, and $\gamma = 10dB$.
2: **Initialize:** $\boldsymbol{\theta}$ and $\boldsymbol{\beta}$.
3: **for** Training epoch = 1 to $\varepsilon$ **do**
4:     Sample a batch of dataset $\{\mathbf{S}_1, \mathbf{S}_2, ..., \mathbf{S}_B\} \in \mathcal{D}$;
5:     **for** Input image $\mathbf{S}_b$ with $b = 1$ to $B$ **do**
6:         Encode $\mathbf{S}_b$ by $f_{\text{E}}(\mathbf{S}_b)$ with masking ratio $\eta$;
7:         Encode $\mathbf{S}_b$ by $f_{\text{E}}^{\text{Ieal}}(\mathbf{S}_b)$;
8:         Calculate the sample of the semantic noise by $\widetilde{\mathbf{n}}_b = f_{\text{E}}^{\text{Ieal}}(\mathbf{S}_b) - f_{\text{E}}(\mathbf{S}_b)$;
9:         Transmit the encoded symbols to the receiver through noisy channel;
10:        Decode received signal;
11:        Obtain the output;
12:     **end for**
13:     Estimate the various of the semantic noise with maximum likelihood estimation according to the sample $\{\widetilde{\mathbf{n}}_1, \widetilde{\mathbf{n}}_2, ..., \widetilde{\mathbf{n}}_B\}$;
14:     Calculate the loss function based on (18);
15:     Update the parameters of $\boldsymbol{\theta}$ and $\boldsymbol{\beta}$ by using Adam optimizer based on the loss function.
16: **end for**
17: **Output:** $f_{\text{E}}(\cdot)$ and $f_{\text{D}}(\cdot)$.

---

objective (18) and **Algorithm 1** aim to jointly minimize the data distortion. However, the learned and shared tokens are difficult to perfectly fit the distribution of specific source. Thus, the performance-constrained semantic coding exists an inherent fidelity bottleneck even if the channel condition is good. In order to further improve the system performance of the ESemCom, we propose the digital-analog transmission based ESemCom framework.

## III. DIGITAL-ANALOG TRANSMISSION BASED EMERGENCY SEMANTIC COMMUNICATION

In this section, we further propose DA-ESemCom framework to break the inherent fidelity bottleneck from the performance-constrained semantic coding scheme. We adopt traditional DSC scheme to enhance the fidelity of reconstructed image by the semantic coding. Different from traditional source coding, the DSC scheme transmits a few encoded symbols to correct the errors of the reconstructed image rather than participate directly in reconstructing the data. As a result, the distorted encoded symbols of the DSC can enhance the fidelity of the reconstructed image as much as possible even at poor channel conditions. Inspired by DA transmission framework, we integrate the analog semantic coding and the digital DSC schemes to leverage their respective advantages, thus optimizing overall performance [26].

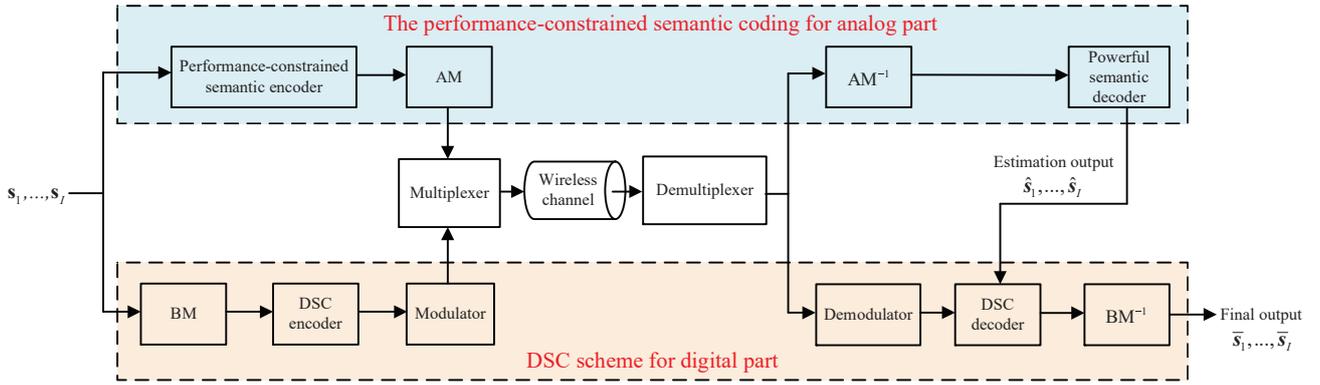

Fig. 2: The proposed DA-ESemCom framework for enhancing the performance-constrained semantic coding transmission.

Figure 2 shows our proposed DA-ESemCom framework, which aims to enhance the performance-constrained semantic coding transmission and consists of the analog and digital parts. In particular, the multiplexer and the demultiplexer are adopted to transmit the channel input signals of both the analog part and the digital part over the same wireless channel [27], [28], thus improving spectrum efficiency. As shown in Fig. 2, the analog part is the proposed performance-constrained semantic coding scheme, which provides robust and efficient wireless transmission. Without loss of generality, the Analog Mapping (AM) achieves by the normalization and reshape layers to map the feature representation vectors into complex-valued channel inputs that satisfy the average transmit power constraint. At the semantic decoder, the another analog mapping ($AM^{-1}$) is responsible for transforming the complex-valued channel outputs into real-valued received signals, and it is achieved by the reshape layer. The received signals are fed into subsequent semantic decoder, thus obtaining an estimation output. In particular, the estimation output will used as the decoding input of the digital part, thus effectively integrating the analog and digital parts to maximize the fidelity of final output.

As shown in Fig. 2, the proposed digital part adopts high-precision Bit Mapping (BM) and DSC schemes for digital transmission. At the transmitter, the BM is responsible for mapping the sources into bit vectors. In practice, when the sources are non-overlapping image patches of an image, the BM consists of the Discrete Cosine Transformation (DCT) and quantization modules to map the pixels to the bits. Then, the DSC encoder adopts digital channel coding to encode the bits and only outputs the parity bits for digital transmission. By using the modulator, the parity bits is modulated into channel input signal. At the receiver, the digital part performs the inverse operation of the transmitter, where the demodulator is responsible for mapping received complex-valued signals back to received parity bits. Then, the DSC decoder adopts the received parity bits and the bits of the estimation output to execute decode. In order to obtain final output data, anther BM ($BM^{-1}$) is used to map the decoded result of the DSC decoder back to the data space of the sources, where $BM^{-1}$ is the inverse operation of BM, namely Inverse Discrete Cosine Transform (IDCT) and dequantization. The final output of the proposed DA-ESemCom framework is defined as $\overline{\mathbf{S}} = \{\overline{\mathbf{s}}_i\}_{i=1}^{I}$.

## IV. DA TRANSMISSION-BASED CODING SCHEME FOR PROPOSED DA-ESEMCOM

In emergency communication scenarios with limited power and bandwidth resources, it is also essential to develop DA transmission-based coding scheme in order to minimize distortion between the source and the final output at the target channel SNR, thereby enhancing the fidelity of the source data. Specifically, we need to consider the trade off between the transmission power and bandwidth of the analog and digital coding parts. For the power allocation, we define the transmission power for the digital part, the transmission power for the analog part, and total transmission power as $P_d$, $P_a$ and $P_t$, respectively. The power allocation constraint for the proposed DA-ESemCom framework can be expressed as follows:

$$P_a + P_d \leq P_t. \quad (19)$$

Furthermore, let be $B_d$, $B_a$ and $B_t$ available bandwidth for the digital part, available bandwidth for the analog part, and total available bandwidth, respectively. The bandwidth allocation for the proposed DA-ESemCom framework is given as follows:

$$B_a + B_d \leq B_t. \quad (20)$$

In the proposed DA-ESemCom framework, the semantic encoding based analog transmission does not require significant power and bandwidth resources for reliable transmission of the sources due to the robustness and effectiveness of the semantic encoding. For the digital DSC encoding and transmission, it involves lower bandwidth consumption for transmitting parity bits. However, it incurs higher power consumption to ensure reliable digital transmission over noisy channels. Therefore, our proposed resource allocation scheme for the DA-ESemCom framework prioritizes the allocation of power and bandwidth for reliable digital transmission, while ensuring that the remaining resources are sufficient to achieve an ideal performance of robust and efficient analog semantic transmission. In particular, the proposed resource allocation scheme avoids complex iterations, rendering it lightweight and efficient. Consequently, it is well-suited for the mobile transmitter with limited hardware and software capacity in emergency communications. In the following, we compute the power and bandwidth to be allocated for the digital transmission and the analog semantic transmission, respectively.

## A. Digital Encoding and Transmission

Let $r_{\text{cod}}$ and $r_{\text{mod}}$ (bits/symbol) be the coding rate of the DSC encoder and the transmission rate of the modulation, respectively. For the DSC encoder, we denote by $N$ the number of information bits within each coded block. After the source to be bit mapping, we obtain the bit vectors of the sources, where total length of the bit vectors is defined as $M$. Thus, we need to adopt $\lceil \frac{M}{N} \rceil$ times of the DSC encoding, where $\lceil \cdot \rceil$ is the ceil function. The DSC encoding and the digital transmission only transmit the parity bits. The number of bits to be transmitted denoted by $C$, is given as follows:

$$C = \left\lceil \frac{M}{N} \right\rceil \times N \times r_{\text{cod}}. \quad (21)$$

Then, $C$ parity bits are combined into transmission blocks for wireless transmission. Considering the Rayleigh slow fading channel, the received block, denoted by $\widetilde{\mathbf{y}}_k \in \mathbb{C}^{V \times 1}$ with $k = 1, 2, ..., K$, can be given as follows:

$$\widetilde{\mathbf{y}}_k = \widetilde{\mathbf{x}}_k h_k + \mathbf{n}, \quad (22)$$

where $\widetilde{\mathbf{x}}_k \in \mathbb{C}^{V \times 1}$ with $k = 1, 2, ..., K$ denotes transmission block, $K$ represents total number of transmission blocks, each transmission block is with $V$ complex-value symbols, and $h_k$ is the channel coefficient. It is fact that one complex-value symbol can carry two real-valued instances of the semantic feature vector. Thus, total number of transmission blocks $K$ is given by

$$K = \left\lceil \frac{C \times r_{\text{mod}}}{2V} \right\rceil. \quad (23)$$

Then, the bandwidth consumption of the digital transmission can be given as follows:

$$B_{\text{d}} = KV. \quad (24)$$

In order to ensure error-free demodulation of the digital symbols at varying channel, we need to allocate transmission power in terms of the error probability $\varepsilon$ and the modulation mode $\phi$, where $\phi$ is corresponding to $r_{\text{mod}}$. For example, we adopt Quadrature Phase-Shift Keying (QPSK), namely, $r_{\text{mod}} = \frac{1}{4}$ and $\phi = 4$. We denote by $b_{\text{avg}}$ the average energy for each transmission bit. Then, $\varepsilon$ can be expressed as follows [29]:

$$\varepsilon = 2\left(1 - \frac{1}{\phi}\right) f_{\text{GQ}}\left(\sqrt{\frac{6 \log \phi}{\phi^2 - 1} \times \frac{b_{\text{avg}}}{n}}\right), \quad (25)$$

where $n$ is the AWGN power and the Gaussian Q-function $f_{\text{GQ}}(\cdot)$ can be expressed by

$$f_{\text{GQ}}(x) = \int_x^\infty \frac{1}{\sqrt{2\pi}} \exp\left(-\frac{x_2}{2}\right) dx. \quad (26)$$

Then, we can obtain the following expression regarding the average energy for each transmission bit as follows:

$$b_{\text{avg}} = \frac{f_{\text{GQ}}^{-1}\left(\frac{\varepsilon}{2(1-\frac{1}{\phi})}\right) \times n}{\sqrt{\frac{6 \log \phi}{\phi^2 - 1}}}, \quad (27)$$

where $f_{\text{GQ}}^{-1}(\cdot)$ is the inverse of (26). Although the digital transmission is less robust than the analog transmission, it can achieve an intended threshold of the error probability $\varepsilon_{\text{th}}$ by allocating enough transmission power. Thus, the transmission power consumption for the digital part $P_{\text{d}}$ with an error probability threshold $\varepsilon_{\text{th}}$ is given by

$$P_{\text{d}} = C \times \frac{f_{\text{GQ}}^{-1}\left(\frac{\varepsilon_{\text{th}}}{2(1-\frac{1}{\phi})}\right) \times n}{\sqrt{\frac{6 \log \phi}{\phi^2 - 1}}}, \quad (28)$$

where we can obtain the value of $f_{\text{GQ}}^{-1}(\cdot)$ under both the error probability threshold $\varepsilon_{\text{th}}$ and the modulation mode $\phi$ by simply looking the Gaussian Q-function table.

## B. Analog Encoding and Transmission

In order to fully use the limited bandwidth and power resources, we let $B_{\text{a}} = B_{\text{t}} - B_{\text{d}}$ and $P_{\text{a}} = P_{\text{t}} - P_{\text{d}}$ based on (20) and (19). With the DL-based semantic encoder, the semantic feature extracted from the source is directly mapped to the complex-valued channel input for wireless transmission. To fully use remaining bandwidth $B_{\text{a}}$, the number of sources for analog encoding and transmission based on (1) can be given as follows:

$$Q = \left\lfloor \frac{2B_{\text{a}}}{L} \right\rfloor, \quad (29)$$

where $\lfloor \cdot \rfloor$ represents the floor function. For the analog transmission, we consider to scale the output of the non-linear compression by using the scaling factors before adopting the analog mapping. In particular, the scaling factor satisfies the power constraint of the analog part $P_{\text{a}}$ and is inversely proportional to the standard deviation of the sources to minimize the distortion caused by the channel noise [A]. The scaling factor, denoted by $g_i$ with $i = 1, 2, ..., Q$, is given as follows:

$$g_i = \sqrt{\frac{P_{\text{a}}}{\sigma_t \sum_{j=1}^Q \sigma_j}}. \quad (30)$$

Then, the normalized channel input of the the analog part is given as follows:

$$\mathbf{x}_i = \sqrt{L}g_t \frac{\widetilde{\mathbf{x}}_i}{\sqrt{\widetilde{\mathbf{x}}_i^\dagger \widetilde{\mathbf{x}}_i}} \quad (31)$$

where $\widetilde{\mathbf{x}}_i$ is the last convolutional layer of the semantic encoder and $\widetilde{\mathbf{x}}_i^\dagger$ represents the conjugate transpose of $\widetilde{\mathbf{x}}_i$. Then, we develop the digital-analog encoding scheme for the proposed DA-ESemCom framework as depicted in **Algorithm 2**.

**Algorithm 2** Digital-Analog Encoding Scheme

1: **Input:** $\mathbf{S}$, $B_t$, $P_t$, $r_{cod}$, $r_{mod}$, $\phi$, and $\varepsilon_{th}$.
2: **Initialize:** $B_d = P_d = B_a = P_a = 0$.
3: Compute $\boldsymbol{\rho}^2 = \{\rho_1^2, \rho_2^2, ..., \rho_I^2\}$ based on (15);
4: Compute $B_d$ based on (21), (23), and (24);
5: Compute $P_d$ based on (28);
6: Digital encode $\mathbf{S}$ according to the parameters of $r_{cod}$ and $r_{mod}$;
7: Obtain $B_a = B_t - B_d$;
8: Obtain $P_a = P_t - P_d$;
9: Compute $Q$ based on (29);
10: Select $Q$ sources $\mathbf{S}_T = \{\mathbf{s}_1, \mathbf{s}_2, ..., \mathbf{s}_Q\}$ with low value of $\rho_i^2 \in \boldsymbol{\rho}^2$ from $\mathbf{S}$;
11: The rest of sources $\mathbf{S}_D = \{\mathbf{s}_{Q+1}, \mathbf{s}_{Q+2}, ..., \mathbf{s}_I\}$ are dropped;
12: Compute $\mathbf{g} = \{g_1, g_2, ..., g_Q\}$ of $\mathbf{S}_T$ based on (30);
13: Encode $\mathbf{S}_T$ based on (31);
14: **Output:** $\widetilde{\mathbf{X}} = \{\widetilde{\mathbf{x}}_1, \widetilde{\mathbf{x}}_2, ..., \widetilde{\mathbf{x}}_K\}$ and $\mathbf{X} = \{\mathbf{x}_1, \mathbf{x}_2, ..., \mathbf{x}_Q\}$.

## V. NUMERICAL RESULTS

### A. Simulations Setup

Fire is a frequent disaster, which can cause significant damage to lives and property within a very short time. In the context of such a disaster, the proposed DA-ESemCom framework is adopted to encode and forward the scene images for the human decision-making and AI-driven method, where the digital part adopts 3/4 Low-Density Parity Check Code (LDPC) [30] for the DSC scheme and QPSK for the modulation. Specifically, we attempt to analyze the disaster environment according to recovered image. Apart from the human decision-making, the object detection is considered in this simulation, where pre-trained YOLO [31] is adopted to obtain the semantic label of the image in the object detection. Then, we adopt classical Peak Signal to Noise Ratio (PSNR), widely used Multi Scale Structural Similarity Index Measure (MS-SSIM) [32], and DL-based Learned Perceptual Image Patch Similarity (LPIPS) [33] metrics to measure data fidelity. For measuring detection performance, the mean Average Precision (mAP) [34] is adopted to estimate the effectiveness of reconstructed image serving for AI-driven object detection. Except for the objective evaluation, we also analyse some visual examples of the reconstructed images for the object detection in terms of different transmission schemes. For the simulation parameters, we consider the IEEE 802.11 standards for the applications of low-power and low-latency, the available bandwidth $B$ is set to 20MHz [35]. The total power is set to 1W for a wide range of surveillance applications at the EWC devices.

### B. The Adopted Datasets

In order to train the proposed DA-ESemCom framework as wall as measure its performance, we employ two widely used datasets:

**ImageNet-1K** [36] contains 1000 different categories in the training set and each category contains about 1000 images. The dataset is widely used for various AI-driven vision tasks, such as classification, detection, and segmentation. We use this dataset to pre-train our proposed semantic coding, laying the basis for learning the optimal coded representations.

**COCO2017** [37] contains 7,157 images and includes different kinds of fires such as building, wildland and residential fire, containing shots captured at day time, dusk or night time. This diversity is convenient to evaluate the performance of the system under different fire scenarios.

### C. The Adopted Comparison Schemes

We compare our proposed DA-ESemCom framework with the classical Separated Source-Channel Coding (SSCC), the ideal semantic transmission, and two DL-based JSCC schemes. For the proposed DA-ESemCom framework and two DL-based JSCC schemes, the extracted semantic features are mapped into the 4-bits constellation points as the channel input, which makes it possible to implement in the existing RF system of EWC. The details of the comparison schemes are given as follows:

*1) Classical SSCC:* The classical SSCC is based on Shannon's reliable transmission framework, where widely used Joint Photographic Experts Group (JPEG) [38] and LDPC are adopted as the source coding and the channel coding, respectively.

*2) Ideal semantic transmission:* The ideal semantic transmission scheme, serving as the performance upper bound of the proposed semantic coding scheme, is with unlimited channel resources, where full, noiseless, and high-dimensional semantic feature vectors can be transmitted to the receiver with full-resolution constellation.

*3) DL-based JSCC:* The DL-based JSCC schemes adopt the proposed semantic codec without using the DSC part, where DL-based JSCC (Proposed loss) and DL-based JSCC (MSE loss) schemes are trained with the proposed optimization objective (18) and only MSE objective (17), respectively.

### D. Objective Quality for Different Methods

Figure 3 shows the performance of reconstruction results achieved by various schemes versus channel SNR over fading channel. It can be seen from Fig. 3(a) that the PSNR achieved by proposed DA-ESemCom framework outperforms two DL-based JSCC and classical SSCC schemes in all channel conditions. In particular, two DL-based schemes fail to reach the baseline of ideal semantic transmission in all channel conditions, whereas the proposed DA-ESemCom framework can achieve better PSNR than the ideal semantic transmission when the SNR exceeds 2 dB. It proves that by the introduction of the DA transmission framework into the semantic transmission scheme can break its performance bottleneck in terms of data fidelity. The DL-based JSCC (Proposed loss) scheme achieves superior PSNR values than the DL-based JSCC (MSE loss) scheme. The increased performance of the DL-based JSCC (proposed loss) scheme is attributed to further reduce the distortion caused by the semantic noise and channel noise, as guided by the proposed overall distortion loss. For the classical SSCC scheme, it fails to approach the other baselines even at a channel SNR of 20 dB. The low PSNR values of the SSCC scheme is attributed to the decoding failed caused by severe distortion of received symbols.

Apart from the pixel-wise PSNR metric, regarding the performance on preserving semantic information, Fig. 3(b) shows the LPIPS values versus the channel SNRs over fading channel. Since the finite constellation mapping and synonymous mapping error inevitably lead to the loss of semantic information, the DL-based JSCC schemes have poorer performance on LPIPS than the baseline of the ideal semantic transmission. Also, the LPIPS values achieved by classical SSCC scheme is close to the DL-based JSCC schemes (MSE loss) when the SNR exceeds 16 dB. The DL-based JSCC schemes have good robustness at the low SNR region, but experience performance saturation at high SNR region. By integrating the DL-based JSCC and DSC schemes, our proposed DA-ESemCom framework has the advantages of robust data recovery and strong data fidelity. Thus, the LPIPS values achieved by our proposed DA-ESemCom framework obviously overpasses other schemes.

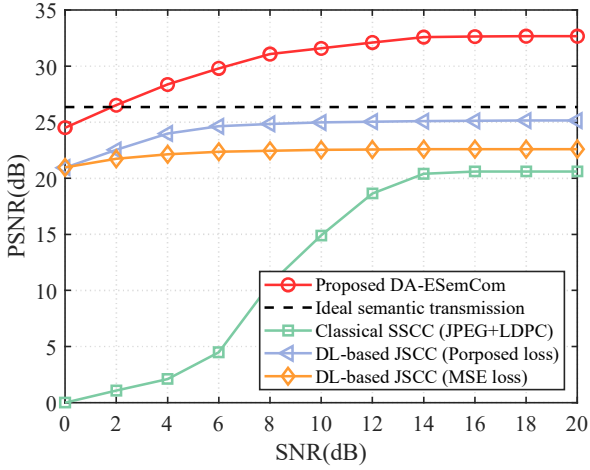
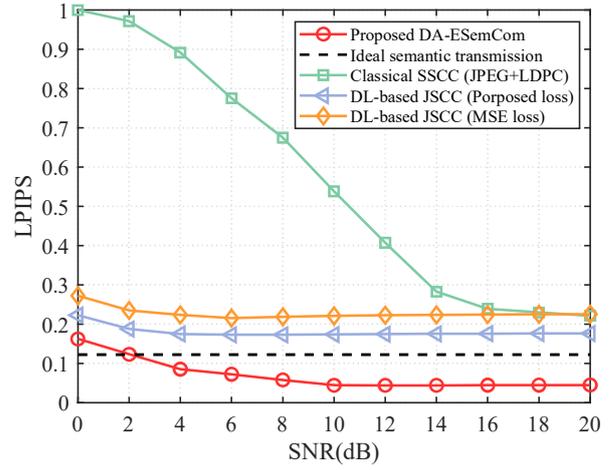

Fig. 3: Performance of reconstruction results of the proposed DA-ESemCom framework and the other baselines over fading channel.

Figure 4 shows the performance on the mAP metric of various schemes versus channel SNR over fading channel. In particular, the mAP metric usually includes mAP50 and mAP50:95 scores, which adopt low and high detection thresholds to calculate their mAP scores, respectively. Compared with the mAP50 score, the mAP50:95 score is used to evaluate the object detection application with strict accuracy requirement. Obviously, the mAP50 and mAP50:95 scores achieved by the proposed DA-ESemCom framework outperform other schemes at all channel SNRs and closely resemble the original images at the high SNR region. For the EWC network, the reconstructed images of proposed DA-ESemCom framework meet the requirement of various surveillance applications. In addition, the DL-based JSCC schemes can achieve graceful degradation in terms of the mAP50 and mAP50:95 scores due to its strong noise robustness. By contrast, the classical SSCC scheme experiences cliff effect, which results in a significant reduction of the mAP when the channel SNR falls beneath the designed SNR. However, the mAP achieved by the classical SSCC scheme has a sharp improvement when the SNR exceeds 8 dB. The increased performance of the classical SSCC scheme is attributed to the ability strong data fidelity but relies on good channel condition and sufficient transmission bandwidth. The idea of the proposed DA-ESemCom framework is that the DL-based JSCC can restore a basic data without spending too much transmission resources. Then, the DSC scheme can significantly enhance the data fidelity of the basic data. By joint resource optimization, the proposed DA-ESemCom framework can obtain the optimal performance.

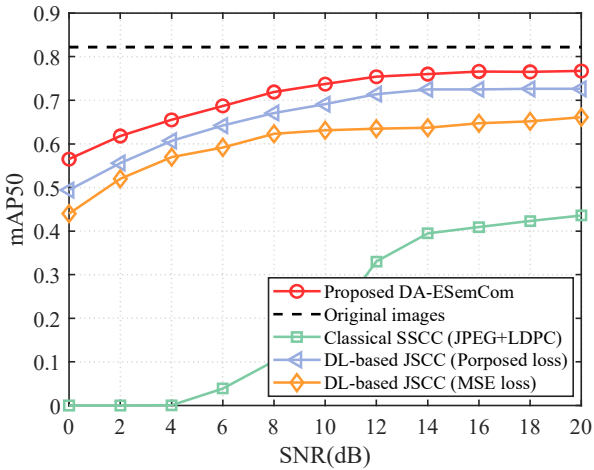
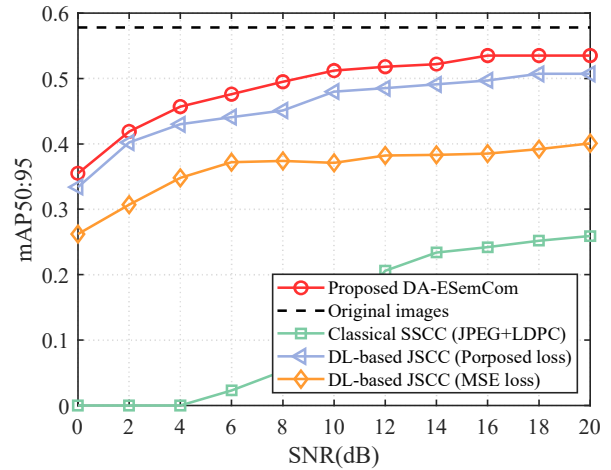

Fig. 4: Performance on object detection of the proposed DA-ESemCom framework and the other baselines over fading channel.

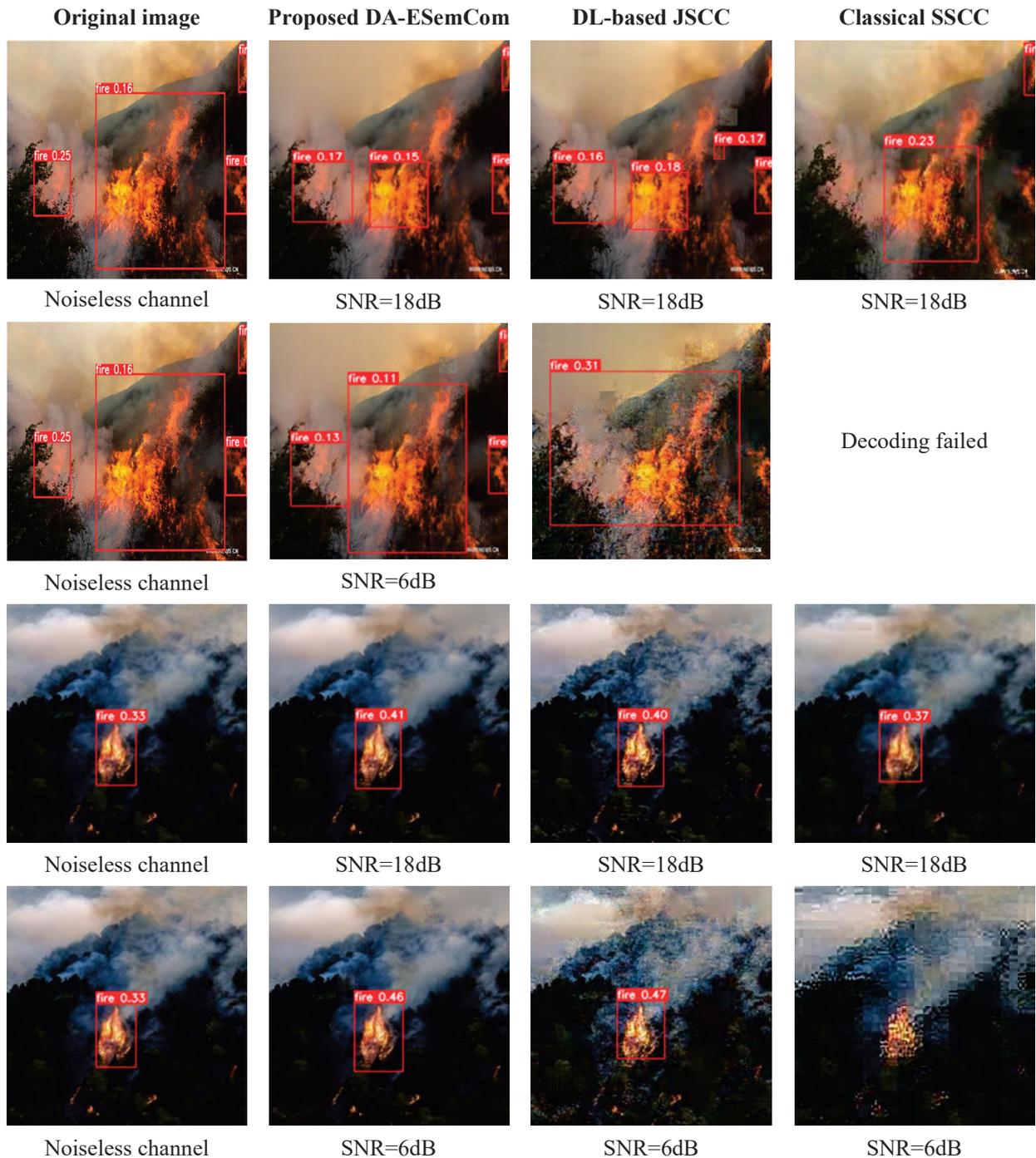

Fig. 5: Visual examples of object detection results on the reconstructed images of the proposed DA-ESemCom framework and the other baselines under different channel conditions.

*E. Subjective Quality for Different Methods*

More intuitively, Fig. 5 shows the visual examples of object detection results on the reconstructed images of the proposed DA-ESemCom framework and the other baselines under different channel conditions. We randomly select two reconstructed images as visual examples for the fire detection. From the results, the reconstructed images of the DL-based JSCC and classical SSCC schemes fail to obtain the detection results close to the original images due to the blocking artifacts at all channel conditions. In particular, the reconstructed images of classical SSCC scheme focus on keeping the pixel-wise consistency with original image but are highly susceptible to serious distortion or decoding failed due to channel noise. Due to the strong noise robustness, the DL-based JSCC scheme can reconstruct a image for detection. Nevertheless, the semantic coding employed in DL-based JSCC inherently suffers from a fidelity bottleneck, resulting in suboptimal reconstruction quality and reduced detection performance in certain scenarios. Our proposed DA-ESemCom framework generates visually satisfactory reconstructed images across all considered channel SNRs. Furthermore, the fires can be correctly localized, but are with a lower confidence score than original images.

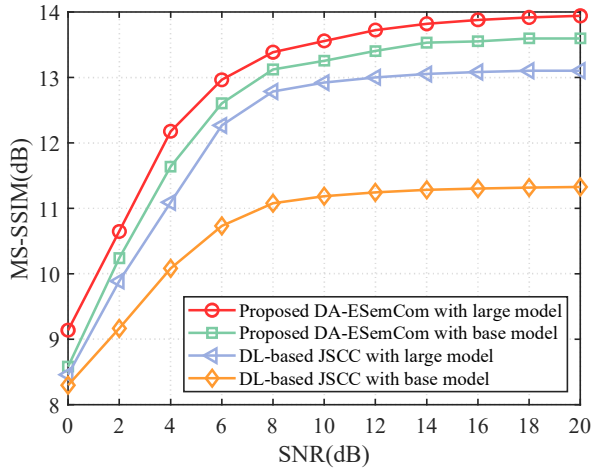 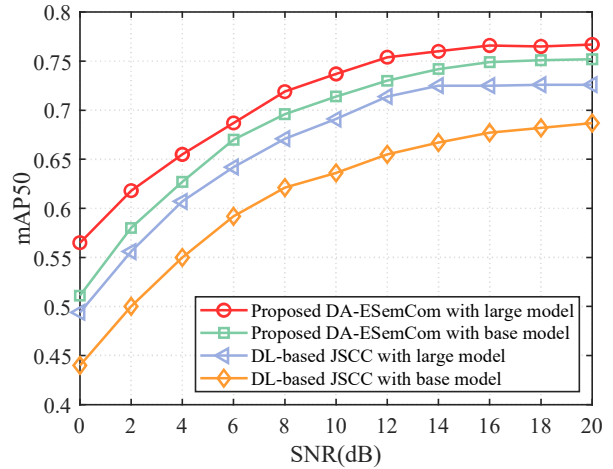

(a) MS-SSIM versus the channel SNRs  (b) mAP50 score versus the channel SNRs

Fig. 6: Ablation study results in terms of the PSNR and the mAP50 performances over the fading channel.

TABLE I: The neural network parameters of the semantic encoder.

| Model size \ Model setting | Transformer blocks | Embedding dimension | Depth | Heads | Output dimension |
|---|---|---|---|---|---|
| Large | 8 | 1024 | 24 | 16 | 512 |
| Base | 6 | 768 | 12 | 12 | 512 |

*F. Objective Quality for Ablation Study*

In the following, we develop an ablation study to verify the effectiveness of the proposed DA-ESemCom for device with different computing resources. For the above proposed DA-ESemCom framework and the DL-based JSCC scheme, we have proposed using the Transformers backbone with different model sizes to construct the proposed semantic encoder. Table I shows the parameters of the semantic encoder with base model or large model.

Fig. 6 also shows the ablation study results in terms of the MS-SSIM and the mAP50 performances over the fading channel. It can be seen from Fig. 6(a) that the MS-SSIM values achieved by the DL-based JSCC with large model scheme outperform those of the base model by approximately 2.7 dB when the SNR exceeds 8 dB. With the proposed DA-ESemCom framework, the MS-SSIM values surpassing those of the DL-based JSCC large model by approximately 0.3 dB. This illustrates that the proposed DA-ESemCom framework can be applied efficiently in lightweight device and provides better fidelity performance without using large NN model. Also, the proposed DA-ESemCom framework can enhance the MS-SSIM performance of the large model by an additional 0.3 dB. It also can be seen from Fig. 6(b) that the proposed DA-ESemCom can effectively improve the mAP50 performance of the base model and large model, respectively. With the proposed DA-ESemCom framework, the reconstructed images of the base model can provide better detection performance than the large model. Overall, our proposed DA-ESemCom framework can improve effectively the performance of the DL-based JSCC scheme regardless of the deployed model size.

VI. CONCLUSION

In this paper, we investigated the semantic communication framework, which is deployed to the mobile devices in the UDP-based EWC networks. Considering the EWC environment, we proposed performance-constrained semantic coding model, which considers the effects of channel noise as well as semantic noise for semantic transmission. Then, we derived Cramér-Rao lower bound of the proposed semantic coding model. It has been shown that the performance-constrained semantic codec exists an inherent performance bottleneck. Thus, the performance-constrained semantic coding scheme cannot achieve the ideal performance of its design in practice. To improve the system performance of ESemCom, we have proposed a novel DA-ESemCom framework and DA transmission-based coding scheme. We showed that our proposed method outperforms classical SSCC and DL-based JSCC schemes under the same channel conditions. The proposed DA-ESemCom framework and other schemes were simulated under fading channel with varying SNR values. Performance metrics such as PSNR, mAP, and MS-SSIM were used to assess its effectiveness. The simulation results have validated that the proposed DA-ESemCom framework performs better than the classical SSCC and DL-based JSCC schemes in terms of reconstruction quality and detection performance. Also, the ablation study has shown that the proposed DA-ESemCom framework can improve effectively the performance of the DL-based JSCC scheme regardless of the deployed model size.